\documentclass[preprint,10pt]{amsart}

\usepackage{amsmath}
\usepackage{amssymb} 
\usepackage{booktabs} 
\usepackage{amssymb} 
\usepackage{amsthm}
\usepackage{caption}
\usepackage{booktabs} 
\usepackage{graphicx} 
\usepackage{amsmath} \usepackage{courier} \usepackage{color} \usepackage{epsfig}
\usepackage{xcolor} \usepackage{colortbl,booktabs} \usepackage{xspace}
\usepackage{capt-of}
\usepackage{mathtools}
\usepackage{pgffor}

\usepackage[obeyspaces]{url} \usepackage{hyperref} \hypersetup{colorlinks=true,
	urlcolor=blue}

\usepackage{subcaption}
\usepackage[figuresright]{rotating} 
\newcommand{\co}{CO$_2$\xspace}

\newcommand{\rmd}{\mathrm{d}}
\newcommand{\diag}{\mathrm{diag}}
\renewcommand{\vec}[1]{\textbf{#1}}
\def \casename {1}
\def \casenamec {-2}
\def \ffac {0.9}
\begin{document}		
		
\title[]{USING FLOW MODELS WITH SENSITIVITIES TO STUDY COST EFFICIENT MONITORING
  PROGRAMS OF \co STORAGE SITES}

\author{Halvor M\o ll Nilsen}
\author{Stein Krogstad}
\author{Odd Andersen}
\author{Rebecca Allen}
\author{Knut-Andreas Lie}

\address{SINTEF DIGITAL, Mathematics and Cybernetics, PB 124 Blindern, N-0341 Oslo,
  Norway}

\begin{abstract} 
  A key part of planning CO2 storage sites is to devise a monitoring strategy. The
  aim of this strategy is to fulfill the requirements of legislations and lower
  cost of the operation by avoiding operational problems. If CCS is going to be a
  widespread technology to deliver energy without CO2 emissions, cost-efficient
  monitoring programs will be a key to reduce the storage costs.  A simulation
  framework, previously used to estimate flow parameters at Sleipner Layer 9
  [1], is here extended and employed to identify how the number of
  measurements can be reduced without significantly reducing the obtained
  information. The main part of the methodology is based on well-proven, stable
  and robust, simulation technology together with adjoint-based sensitivities and
  data mining techniques using singular value decomposition (SVD). In particular
  we combine the simulation framework with time-dependent (seismic)
  measurements of the migrating plume. We also study how uplift data and
  gravitational data give complementary information.
  
  We apply this methodology to the Sleipner project, which provides the most
  extensive data for CO2 storage to date. Since injection commenced in 1996, a
  series of seismic images have been taken to capture the migrating CO2 plume. 
  First, we obtain a direct match with current measurements by calibrating
  topography, permeability, CO2 density, porosity, and injection rates. Using an
  estimate of the misfit, we show how one can minimize the number of
  measurements without significantly influencing the accuracy of the
  parameter estimates. We also investigate how assumptions on measurement
  errors influence the parameter estimate uncertainties.  
  
  The original benchmark model of Sleipner Layer 9  assumed a CO2 density of 760
  kg/m3 and a significantly lower permeability than what we obtain from the above
  estimate. Using this as the assumed physical situation we show how the
  measurement sensitivities depend on the main dynamics of the systems. In
  particular we show that the sensitivity to the topography near the injection
  point is significantly less than for the matched case. 
  
  We also study the effect of measurement sparsity in space and time and
  show that for the current description of Layer 9, a limited number of
  measurements of the CO2-Water contact is sufficient to estimate the main
  dynamics The results are robust with respect to the choice of measurement
  points, despite the fact that the dynamics is sensitive to small changes in the
  top-surface topography at particular points, which is identified as spill points. By using a singular
  value decomposition (SVD), we show that the response of a small change in topography at the
  spill points gives global effect for this physical situation. 
  
  Finally, we extend the modelling framework with possibilities to incorporate  
  gravity data based on gravitational changes. Using the sensitivities, we discuss
  in which situation different measurements can be utilized to give new
  information about the physical model. 
  
  For this study we utilize a vertical-equilibrium (VE) flow model for
  computational efficiency as implemented in the open-source software MRST-co2lab.
  However, our methodology for deriving efficient monitoring schemes is not
  restricted to VE-type flow models, and at the end, we discuss how the
  methodology can be used in the context of full 3D simulations.
  
\end{abstract}

\maketitle
\section{Introduction}
\label{main}

The Norwegian Continental Shelf offers enormous volumes of potential storage
capacity for
\co in saline aquifers \cite{CO2atlas:NorthSea}. To enable large-scale \co
storage,
prospective operators need effective monitoring strategies for detecting
potential
leakages and other undesired effects such as uplift and subsidence. This will
require
monitoring technologies, whose raw
observation
data should be assimilated into forecast models to confirm that the \co plume is
behaving
as expected and provide support for decision making and remediation planning in
the case
of unpredicted events. We believe that the starting point for designing combined
monitoring--forecast
strategies should be simple robust reservoir models and inversion strategies
localized as much as possible near
the storage aquifer. Depth-integrated flow
models that rely on an assumption of vertical equilibrium have been shown to work
well for aquifers with good-quality sand, which is the most likely candidates
for storage. In this work, we will use this as our reservoir model. The most used
measurement for inversion is based on seismics. Traditionally
seismic is used to estimate the static properties of the full formation, 
but with time-lapse seismic
changes can be estimated. Most processing methods need a detailed model of the overburden. However Marchenko imaging have newly been applied to real field data  \cite{Ravasi2016} and only need limited information of the overburen. A second step, which often makes combining flow modeling and inversion methods difficult is that the estimated quantities do not coincide with the model parameters used in forward simulation. To bridge this gap it is essential introduce model of how changes in pressure and fluid saturations and compositions affect the seismic responses to be able to invert time-lapse seismic to update petrophysical properties of the geological model \cite{Romdhane2014,Dupuy2016b}. We will not consider detailed inversion models for seismic but assume that plume heights can be estimated. In addition, we will use direct gravity changes as a second observation.  

Traditionally, inversion is utilized to
constrain the aquifer modeling. That is, you acquire time-lapse data by example
seismic, gravity or electro magnetic techniques and these data can be inverted to determine changes
in reservoir properties. Then, parameters in the  reservoir model are changed to
minimize the misfit between observed and simulated responses. However, it
is also possible to use the flow model to constrain the seismic inversion so
that the form of the inverted interfaces can be explained by the flow physics.
In fact this is the simples way of ensuring the saturation and pressure changes has
the right physical behavior.  Likewise, one can use the flow model to
investigate the measurability of model parameters, determine the optimal design of
measurements and assess
the effect of acquiring new measurements
(value of information).

To illustrate we use the benchmark model of of the sleipner injection 
\cite{Singh2010}.
In a previous paper \cite{Nilsen2017} we used a simulation model based on vertical-equilibrium to match to the real estimated plume shape \cite{Furre2014,Chadwick2010}.
We use this work as a starting point to investigate
how the sensitivity of the model changes with regard to the physical situation.
To demonstrate this we use the original benchmark model and a member of the family of matched
models. We investigate in detail how these models have different response in
terms of measured quantities like plume-shape and gravity field. In particular,
we show that combinations of gravity data and total mass injected is sufficient
to determine the global parameters. The plume data however give more information
about the top surface in particular for the physical situation of the match
data.

Starting from this, we investigate the most important data in the setting of
optimal design of experiment. 
Rather than a full
investigation in terms of the actual measurements we calculate the optimal design
with respect to the estimate of the plume heights and the gravity changes.
This will give a starting point for evaluating cost-efficient 
monitoring strategies. To this end, we investigate different formulation of
optimal design algorithms which can be extended to a more detailed analysis of the best monitoring strategy. We discuss the importance of having accurate estimates of correlations both in time and space of the estimated quantities to be able to make accurate estimates of the model and to evaluate the best monitoring strategy.

\section{Methods}

\subsection{Governing equations -- Darcy-based approach}\label{sec:gov_equations}
\label{sec:darcy_equations}
The basic mechanisms of \co injection and migration can be modeled using two
components,
water ($w$) and \co ($g$), which at aquifer conditions will appear in an aqueous
and a
supercritical (liquid) phase, respectively. The flow dynamics is described by
mass
conservation,
\begin{align}
\frac{\partial (\phi \rho_{\alpha} s_{\alpha})}{\partial t} + \nabla \cdot
\rho_{\alpha} \textbf{u}_{\alpha} = \rho_{\alpha} q_{\alpha}, \quad \quad
\alpha = w, g,
\label{eqn:MassConservation}
\end{align}
where $\phi$ is the porosity, $s_{\alpha}$ is the fluid saturation,
$\rho_{\alpha}$ the
fluid density, $\textbf{u}_{\alpha}$ the fluid velocity, $q_{\alpha}$ the
volumetric flux
caused by any source or sink, and $\alpha$ denotes the fluid phase. The fluid
velocity is
given by Darcy's equation,
\begin{align}
\textbf{u}_{\alpha} = - k \lambda_{\alpha} (\nabla p_{\alpha} - \rho_{\alpha}
\textbf{g}),
\label{eqn:Darcy}
\end{align}
where $k$ is the absolute rock permeability; $\lambda_{\alpha}=k_{r \alpha} /
\mu_{\alpha}$ is the fluid mobility, where $k_{r \alpha} = k_{r \alpha} (s_w)$
and
$\mu_{\alpha}$ denote the relative permeability and fluid viscosity,
respectively;
$p_{\alpha}$ is the pressure; and $\textbf{g}$ is gravitational acceleration.
The sum of
the saturations are equal to unity, and fluid mobilities and
capillary pressure
($p_c$) are expressed as functions of water saturation,
\begin{gather}
s_w + s_g = 1, \\
\lambda_{\alpha} = \lambda_{\alpha} (s_w), \\
p_c = p_n - p_w = P_c(s_w).
\label{eqn:CapillaryPressure}
\end{gather}
In \cite{Nilsen2017} we demonstrated that for this equation with regard to plume
shape there exist one exact invariant with respect to plume shape given by 
\begin{equation}
\phi = c \bar{\phi}, \quad k=c \bar{k}, \quad q_\alpha = c \bar{q}_{\alpha},
\label{eq:invariant-1}
\end{equation}
where $c$ is a positive constan.,
In addition, there will be an approximate invariant given by $\delta \rho K= (\rho_w-\rho K)
$ a constant.

If we, however extend to gravity response 
\begin{equation} \label{eq:grav_changes}
\Delta \vec{g}(\vec{r}) = \int \frac{(\vec{r} -\vec{x})}{|\vec{r}-\vec{x}|^{3}}
(\rho_w-\rho_{co_2})\, s\, \phi \rmd x^3
\end{equation}
we can no longer change the product of $\Delta \rho \phi$ without changing the
response.

The actual calculations presented in this paper will be done using the vertical
equilibrium model also used in \cite{Nilsen2017}. For a more detailed descriptions see
\cite{co2lab:part2}.

\subsection{Adjoint based gradients} 
In this section we consider matching uncertain topography
and model parameters to observed responses through the use of simulation-based
optimization. Let $x^n$ denote the discrete state variables (heights, pressure,
and well
states) at time $t_n$, and let $F_n$ denote the corresponding discrete versions
of equations
\eqref{eqn:MassConservation}--\eqref{eqn:Darcy} such that
\begin{equation}
\label{eqn:SimulationEqs}
F_n(x^{n-1}, x^n) = 0, \; n = 1,2,\ldots, N,
\end{equation}
constitute a full simulation given initial conditions $x^0$. To match simulation
output to
a set of observed quantities, we augment \eqref{eqn:SimulationEqs} with a set of
model
parameters $\theta^n$ such that,
\begin{equation}
\label{eqn:SimulationEqs_theta}
F_n(x^{n-1}, x^n, \theta^n) = 0, \; n = 1,2,\ldots, N.
\end{equation}
The matching procedure consists of obtaining a set of model parameters $\theta$
that
optimize the fit to observed data, i.e., minimizing a misfit function of the
form
\begin{equation}
\label{eq:objective} J = \sum_{n=1}^N J_n(x^n, \theta^n).
\end{equation}
For obtaining gradients of $J$, we solve the adjoint equations for
\eqref{eqn:SimulationEqs_theta}, see \cite{Jansen2011}. The adjoint equations
are given by
\begin{equation}
\label{eq:adjoint1}
\left(\frac{\partial F_N}{\partial x^N}\right)^T \lambda^N =  - \frac{\partial
	J_N}{\partial x^N}
\end{equation}
for the last time-step $N$, and for the previous time steps $n = N-1, \ldots, 1$,
\begin{equation}
\label{eq:adjoint2}
\left(\frac{\partial F_n}{\partial x^n}\right)^T \lambda_n =  - \frac{\partial
	J_n}{\partial x^n} -
\left(\frac{\partial F_{n+1}}{\partial x^n}\right)^T \lambda^{n+1}.
\end{equation}
Once the adjoint equations are solved for the Lagrange multipliers $\lambda^n$,
the
gradient with respect to $\theta^n$ is given as
\begin{equation}
\label{eq:gradient}
\nabla_{\theta^n} J^n= \frac{\partial J^n}{\partial \theta^n} +
\left(\frac{\partial F^n}{\partial
	\theta^n}\right)^T\lambda^n.
\end{equation}
In this work, we consider a set of parameters $\theta$ constant over time, i.e.,
$\theta^n =
\theta, \; n=1,\ldots,N$, such that
\begin{equation}
\label{eq:parameters}
\theta = \{dz, m_q, m_{\rho}, m_k, m_{\phi}\}.
\end{equation}
In the above, $m_q, m_{\rho}, m_k, m_{\phi}$ are scalar multipliers for rate,
\co density,
and homogeneous permeability and porosity, respectively, while $dz$ is the
absolute
adjustment in top-surface depth of dimension equal to the number of grid cells.

\subsection{Invariants in linearized model}\label{sec:invariants}
From the discussion in Section~\ref{sec:gov_equations}, an exact solution
invariant
parameter family \eqref{eq:invariant-1} is generated by the \emph{basis}
\begin{equation}
\hat{\theta}_1 = \{\mathbf{0}, 1, 0, 1, 1\}. 
\label{eq:exact_invar}
\end{equation}
The scaling that gives an invariant saturation in the
incompressible limit, follows from, see \cite{Nilsen2017} for more details
\begin{equation}
k (\rho_w - \rho_{g})=C \quad \rightarrow \quad  \rmd m_k
\left(\rho_w-\rho_g\right) - \rmd
m_{\rho}\, \rho_g = 0.
\end{equation}
This gives the basis vector
\begin{equation}
\hat{\theta}_2 = \{\mathbf{0}, 0, 1, \left(\frac{\rho_w}{\rho_g}-1\right),
0\}. 
\label{eq:approx_invar}
\end{equation}
The two vectors $\hat{\theta}_1$ and $\hat{\theta}_2$ are not orthogonal, but
this
can be obtained by a redefinition of the parameters.
In the case in which the objective function $J$ only depends on saturation
(i.e.,
we only match plume thickness $h$), these vectors span parameter choices that
give
indistinguishable objective function values. When the multipliers are evaluated
relative to a
minimum in $J$, the vectors $\hat{\theta}_1$ and $\hat{\theta}_2$ are
eigenvectors of the
Hessian of $J$ with zero eigenvalues.

The gravity response is only invariant if $\delta \rho \phi$ is constant.
From our definitions we get
\begin{equation}
\phi (\rho_w - \rho_{g})=C \quad \rightarrow \quad  \rmd m_{\phi} (\rho_w -
\rho_{g}) - \rmd m_{\rho} \rho_g  \rho_g = 0.
\end{equation}
Given that we are in the subspace of invariant plume shapes, we will have
invariant gravity response only if we have changes orthogonal to 
\begin{equation}
N_g =  \{\mathbf{0}, 0,  \left(\frac{\rho_w}{\rho_g}-1\right), -1 \}.
\end{equation}
Similarly changes of total mass is given by
\begin{equation}
\rmd M = M\,  \{\mathbf{0}, 1 ,  1, 0, 0 \}\cdot \rmd \theta \label{eq:total_mass}
\end{equation}

	This is not orthogonal to $N_g$ and thus  
	adding this
	constraint give sufficent information to estimate all the global parameters.
	This is since the change in gravity depends on the volume times the density
	differences while the injected mass depend on the volume times density. Thous
	the combination constrains the volume in the reservoir.

\subsection{Parameter estimation}\label{sec:glsq}\label{sec:svd}
Given a nonlinear model $y=F(\theta)$, between the model parameters, $\theta$, and values observed values , $y$,
the calibrated model against measurements $\hat{y}$ is given by
the minimum of the objective function
\begin{equation}
\label{eq:ObjFun_general}
J= \Big(\hat{y}-F(\theta)\Big)^T\Big(\hat{y}-F(\theta)\Big),
\end{equation}
if the measurements are assumed to be uncorrelated. For a linear model
$F=A\theta$, the
above is the traditional least-square estimate, and can be calculated explicitly
\begin{equation}
\label{eq:lsq}
\tilde{\theta} = (A^T A)^{-1} A^T \hat{y}.
\end{equation}
In this case, it is the best linear unbiased estimator (BLUE). This is a special
case of
general least-square estimate, where the assumption is that the model and the
measurement
process can be modeled by
\begin{equation}
y = A \theta + \epsilon \quad E[\epsilon | A]  = 0 \quad Var[\epsilon| A ] =
\Omega.\label{eq:least_square}
\end{equation}
Here, $\epsilon$ is a stochastic variable, $E[\epsilon | A]$ is the conditional
mean of
$\epsilon$ given $A$, and $Var[\epsilon| A ]$ is the conditional variance.  A usefull reformulation introduce
$Q$ by $\Omega=Q^T  \sigma Q$ such that 

\begin{equation}
	y = A \theta + \sigma Q \widetilde{\epsilon} \quad Var[\widetilde{\epsilon} | A ] = I \label{eq:gleast_square}
\end{equation}

Here the dependence of the measurements is separated in dependence of the values represented by $A$ and the statistical dependence described by $F$. For good estimates one either what statistical independence of linearly related values or  dependence of linearly independent quantities. This reformulation is use full for the understanding, but not efficient for computation due to the difficulty of finding $Q$.

The
general
least-square estimator is most easiliy calculated from the formulation of \eqref{eq:least_square} and given by the minimization of the objective
function
\begin{equation}
J= \Big(\hat{y}-F(\theta)\Big)^T \Omega^{-1} \Big(\hat{y}-F(\theta)\Big).
\end{equation}
Assuming the stochastic variables are Gaussian, the above estimate can be
derived in a Bayesian setting the where $\theta$ is the value which is most
probable given the observed data and the model. This interpretation also
suggests a natural regularization in which quadratic changes of the parameters
are
added as penalties. As such, they can naturally be seen as a result of a trivial
predictor and measurement process in the same way with $A$ the identity.

The generalized least-squares estimator only works well when the matrix has full
rank and
is well conditioned. In the two previous sections, we showed how certain
parameter
combinations give invariant  solution with respect to the
plume thickness $h$. The existence of such non-trivial invariants of the solution in the
parameter spaces will lead
to rank deficiency in $A$.  This can be investigated from the full linear model
by
a singular value decomposition (SVD) of $A$
\begin{equation}
A= U S V',
\end{equation}
where $U$ and $V$ are unitary matrices and $S$ is a diagonal matrix with
non-negative real
numbers $\sigma_i$ on the diagonal, referred to as the singular values
of
$A$. We will refer to the columns of $V$ as the parameter singular vectors and
the column
of $U$ as the response singular vectors, when we consider SVD of a sensitivity
matrix,
since they for a given singular value govern the changes in the parameters
corresponding
to a given response. From the standard deviation of the estimate, we see that
high singular values give low standard deviation for the input variables
corresponding to
this singular value. Parameter singular vectors corresponding to small singular
values
indicate combination of parameters that can not be estimated accurately. 

One should note that the SVD decomposition is
sensitive to the definition and scaling of the parameters so care has to be
taken
when variables of different types are considered. On the other hand, the
standard
deviation of the estimate does not depend on scaling, which shows that the
natural scaling
in this given setting is the inverse of the covariance of the measurements. That
is, for
uncorrelated measurements, the right scaling is the standard deviation.
	Thous a scaling depending on a prior distribution would be naturally. Including a prior in the Bayesian setting could be easily done by adding trivial equations to A with a correspoinding correlation given by the matrix  $\Omega$. This is equivalent with thinking of the prior as an uncertain measurement.

    To evaluate the choise of measurements it is important to evaluate the interplay
    between the covariance of the measurements and the linearly independent the measurements are. This is equivalent to
    consider how the genneral singular values depend on the linear independence of the rows of A with regard to the statistical independence described by $Q$ of equation \eqref{eq:gleast_square} \cite{Paige1985}. The standard relation of statistics that standard deviation scales as $1/\sqrt{n}$ where $n$ is the number of independent measurements are easily recognized as equivalently that the singular value scales in the same way as number of rows is repeated. However if the measurements are related this is not the case, which can be equivalently seen if rows of $Q$ are linearly dependent for rows of rows of $A$  equal. Then now better estimate is achieved as expected when the same measurement is included many times. How ever if we have linearly dependent rows of $Q$ and linearly independent rows of $A$ exact knowledge of one the quantity described by differences in rows of $A$ can be achieved. 
    In this work we will only consider diagonal covariance matrix and the analysis will reduce to SVD of $A$. However it is important to keep in mind which observations can be considered independent, i.e. uncorrelated.

\subsection{Our objective function}

The previous section introduced a general form of the objective function
\eqref{eq:ObjFun_general}. To match our simulation to the interpreted plume heights we seek to minimize $J=\sum J_m$ with
\begin{equation}
\label{eq:matchsim}
J_m(h_m) = \sum_{\text{cells}}V (h^m-h_{\text{obs}}^m)^2 + \sum_{\text{cells}}
\alpha dz^2\, V.
\end{equation}
Here, $m=1,\ldots, M$ denote the time-instances of the set of \emph{observed}
heights
$h_{\text{obs}}$ (e.g., \co plume thickness data taken from literature, etc.),
$h$ are the
simulated heights for the given set of parameters $\theta$
\eqref{eq:parameters}, and $V$
is the volume of the aquifer found below each cell in the 2D top surface grid.
We consider the
top-surface elevation more uncertain that the plume thickness $h$, thous we weight
$dz$ less
in our calculation of the misfit by setting $\alpha=1/9$.

For the work presented here we would use only a member of the family of models
which minimize this objective function. We thus also fix the density and the
porosity to get a unique solution. The values used are $ $ for density and for $
$ porosity. 

\subsection{Optimal design of experiment}
A natural question when planning monitoring is which measurements or observations
give most information about the system we are monitoring. This is often named as
 optimal design of experiments. A key question is then to define a
measure for information. Considering a standard linear model
\begin{equation}
y = A \theta + \epsilon
\end{equation}
where each equation defines a possible measurement of the system. We could then
maximize the information in the sense of minimizing the geometric mean of the
stadard deviation of the estimates of $\theta$.  This will lead to the standard
$D-optimal$ criteria. In the discrete setting this will be
\begin{equation}
\begin{split}
&\max \left(\det(A^T \diag(\lambda) A)\right) \\
&\lambda \geq 0 \quad \sum \, \lambda = N, \quad \quad \lambda \in Z \quad
\lambda < N/m.
\end{split}
\end{equation}
The standard definition is that we can choose to repeat the measurements which
also could be seen as extending the system with equations with the equal rows of
$A$, corresponding to $m=1$. In many cases this is not possible since such measurements with necessarily be correlated. We there for use $m>1$
where $m$ corresponds to the minimal number of different measurements. In most cases this
will give $\lambda>0$ approximately $m$ times. In the limit of $N$ small
compared with the number of possible measurements (i.e number of elements of
$\lambda$)  we can let $\lambda$ be real and in this case the system will be
a convex optimization problem \cite{Grant2014}. We will use an implementation
based on, disciplined convex optimization programing CVX \cite{Grant2014}, which
have the advantage of global convergences. Other similar formulation for optimal design is to
	maximizing trace $(A^T \diag(\lambda) A)$, A-optimal, or the minimum of the
	eigen-values, E-optimal, is also possible. In most cases they give similar
	results and we will only use D-optimal in this work. An other formulation will
be to minimize the number of observations given a restriction on the errors. If
maximum error is considered we can formulate the optimization problem analog to
E-optimal by
\begin{equation}
\begin{split}
&\min \sum \lambda ,\quad \lambda >=0\\
& \min \left(\mathrm{svd}( \diag(\lambda) A) \right)> C, \quad  \left(\lambda < m
\right).
\end{split}
\end{equation}
For this we use Ipopt \cite{Waechter2006}, which is an implementation of the
interior point method. This is a general nonlinear optimization method that is
particular efficient for linear optimization problems which can extended to more
complex definitions since it do not depend on convexity assumptions. All though for the general case it is not globally
convergent. For simple implementation we used the matrix expressions for the
derivatives of the singular value decompositions 
\cite{Giles2008,GilesOnline}, for a more stable implementation and a discussion
in the context of optimal design, see \cite{Walter2010}

We also investigated the use of mutual information to calculate the optimum
placement \cite{Krause2008} using the implementation of \cite{Krause2010}. In this
case it is a pure discrete optimization problem similar to the discrete
A-optimum. One seek to minimize the entropy of the prediction at the point of where an
observation is not made. In general solving the discrete optimization problem is hard however the above code exploit that the the given problem is sub-modular which is similar to the continuous case where one exploit convexity.

\subsection{Building a linear model}
Traditionally, computing the full linear response model is very expensive, in
particular
when using full 3D simulation of the plume migration since we need both a high vertical
resolution to resolve the thin plume and high lateral resolution to capture
caprock
topography. Herein, we will use VE type simulations with adjoint capabilities to
make the
forward simulation more computationally tractable for models with reasonable
lateral
resolution.  To investigate a complete linear model between the parameters and
the
plume height, we can use the adjoint method with one objective function for the
position of
interest defined as $J^n_i = h^n_i$, where $h$ is the plume height and $n$ and
$i$
indicate the position in space and time, respectively.  Then, the adjoint method
can get
all derivatives with respect to $\theta$ by one backward simulation.  Also, the
backward
simulations for the different space and time points can be performed in parallel
and can
also reuse the preconditioner or the LU-decomposition of the system matrix, since
the only
differences between the calculations are in the right-hand sides. Additionally,
information
about the response induced by changes in the parameters at different times can
be
stored. The result is a series of linear relationships,
\begin{equation}
\rmd h^n = A^n \rmd\theta.
\end{equation}
We point out that the backward simulation is linear, which makes this simulation
much
faster than the forward simulations. In addition, the computation time is less
dependent
on the particular objective function, which makes load balancing in a parallel
framework
ideal. 
\begin{figure}[tbp]
	\centering
	\includegraphics[width=1\textwidth]{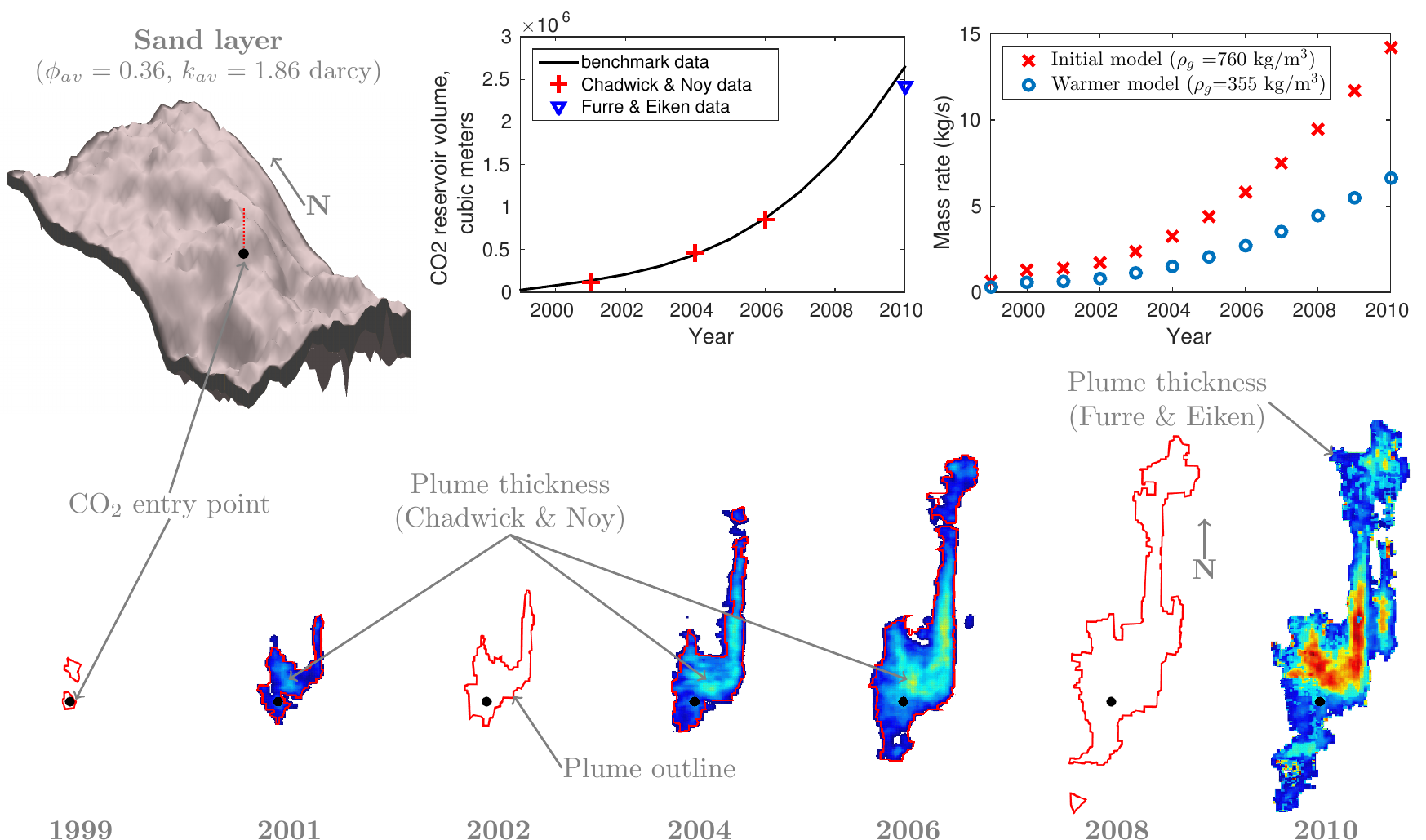}
	\caption{The Sleipner benchmark (sand layer, entry rates, fluid and rock
		properties)
		from \cite{Singh2010}, and \co plume data taken from Chadwick \& Noy
		\cite{Chadwick2010} and Furre \& Eiken \cite{Furre2014}. While seismic
		imaging can
		estimate the volume of \co in Layer~9, the rates
		depends on the
		inferred reservoir density; mass rates shown here are computed assuming
		densities of 760
		and 355 kg/m$^3$.}
	\label{fig:benchmodel}
\end{figure}

\begin{figure}
	\includegraphics[clip=true, trim=210 70 210 10,width=\ffac \textwidth]{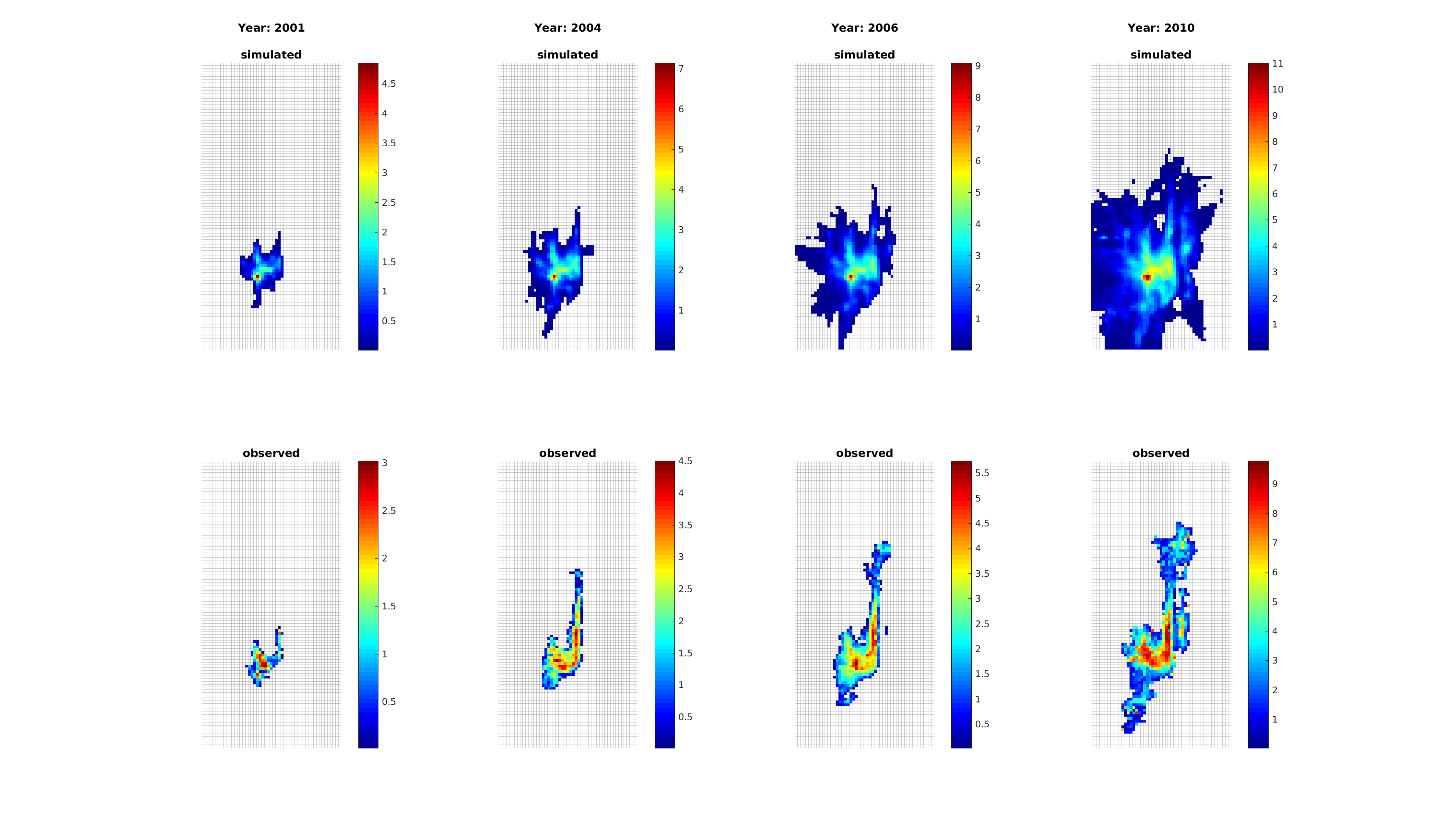}
	\caption{Comparing the original model with the plume data on the simulation grid used.}
	\label{fig:plume_original_data}
\end{figure}
\begin{figure}
	\includegraphics[clip=true, trim=210 70 210 10,width=\ffac \textwidth]{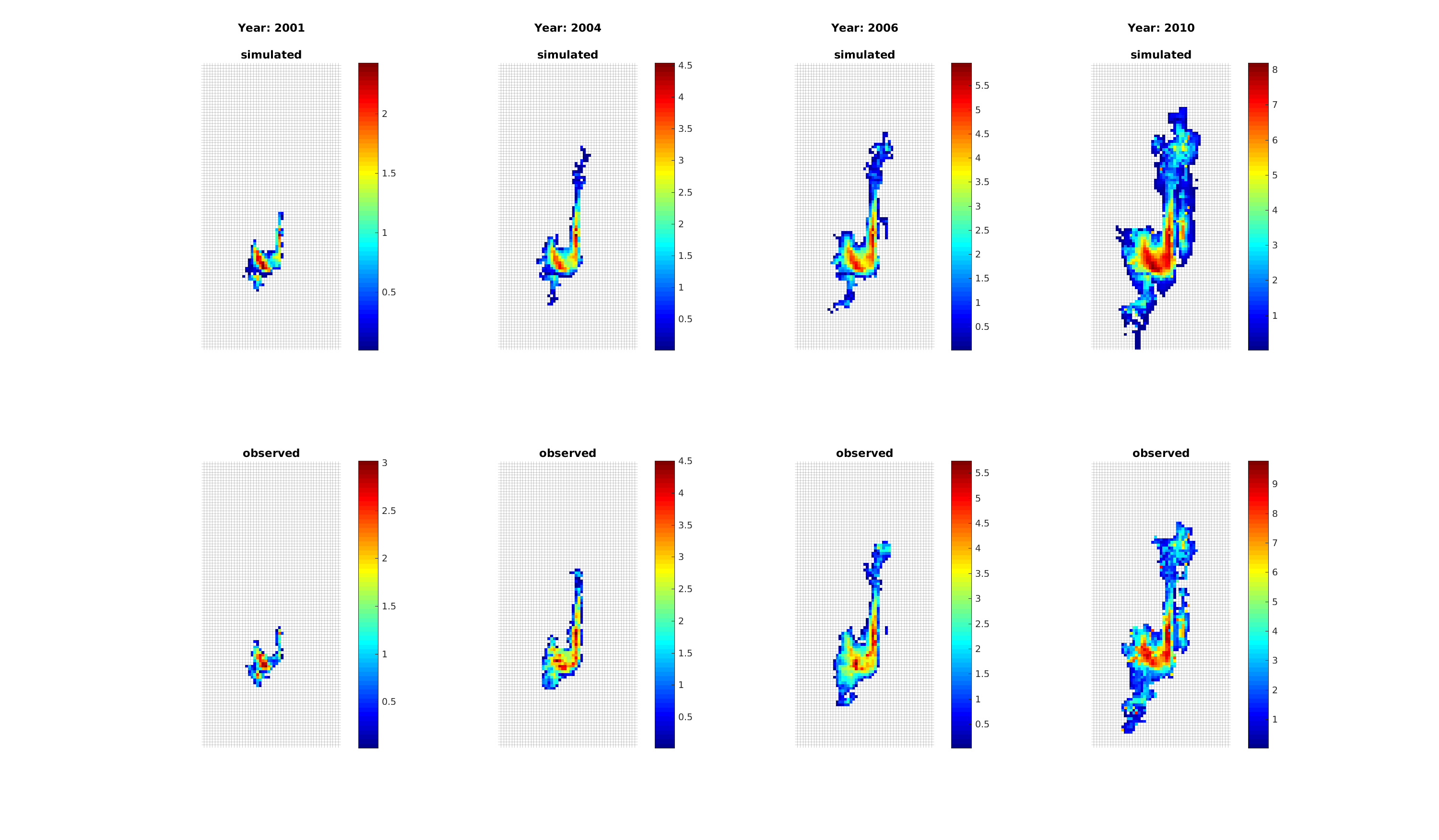}
	\caption{Comparing the matched model with the plume data on the simulation grid used.}
	\label{fig:plume_matched_data}
\end{figure}

\section{The Sleipner Layer~9 (L9) benchmark model}\label{sec:sleipner_model}

We use the Sleipner L9 model based on the benchmark \cite{Singh2010} with the
exact definitions as in our previous work
\cite{Nilsen2017}. Similarly, we use
four sets of observed plume heights $h_{\text{obs}}$ taken from results
published by Chadwick and Noy \cite{Chadwick2010} and Furre and Eiken
\cite{Furre2014}.  From these heights, we 
determined the corresponding plume volumes at four
observation times. An over view of the data is shown in Figure~ \ref{fig:benchmodel}. To
investigate how the response of the model change with different physical
situations we use the original model and a member of the family of match models
investigated in \cite{Nilsen2017}.

To find a unique matched model, we fixed the
density of the model to $478 \mathrm{kg}/\mathrm{m^3}$ and the porosity to $0.37$. The Figure
\ref{fig:plume_matched_data} how the resulting model match the estimated plume
heights form \cite{Chadwick2010,Furre2014}

The most prominent change in the model is that the product of density difference and permeability is changed significantly to achieve a stronger gravity segregation effect. The resulting matched model have a permeability of $13\, \mathrm{Darcy}$ and the volume rate is multiplied with a $0.92$.

\section{Example:}

\subsection{Matching plume on Sleipner}
\label{sec:match_sleipner}
In this paper we will investigate how the physical conditions in the aquifer influence the
sensitivites of the model. We use the original assumption of the Layer 9
benchmark \cite{Singh2010}, as described in section~\ref{sec:sleipner_model}
We now proceed to match our model to the data. We minimize the difference between 
using the same methods as in \cite{Allen2016} given a set of parameters
$\theta$ in the same way as with the linear least-square theory \eqref{eq:lsq}.
The
misfit objective \eqref{eq:matchsim} with observed plume heights
$h_{\text{obs}}$ for the
years 2001, 2004, 2006, and 2010 are used (the first three years of data are
taken from
\cite{Chadwick2010}, and the last year is from \cite{Furre2014}). 
To find a unique matched model, we fixed the
density of the model to $475 \mathrm{kg}/\mathrm{m^3}$ and the porosity to $0.37$. 
The most prominent change in the model is that the product of density difference and permeability is changed significantly to achieve a stronger gravity segregation effect. The resulting matched model have a permeability of $11\, \mathrm{Darcy}$ and the volume rate is multiplied with a $0.91$.
The match before and after the match is shown in Figure~\ref{fig:plume_original_data} and Figure~\ref{fig:plume_matched_data}.

\subsection{Sensitivity to parameter changes}
The sensitivities related to porosity, density, permeability, and rate must be
handled
with particular care since these variables are of different character even if we
have
chosen to define all of them as dimensionless multiplication factors of the
original
values. The starting point for this discussion is to consider the top surface as
constant
so that the changes in \co thickness can be written as
\begin{equation}
\rmd h = A_{m,t}\rmd \theta_m,
\end{equation}
the change in vertical gravity is similarly 
\begin{equation}
\rmd g_z = G_{m,t}\rmd \theta_m.
\end{equation}
In the following we will only consider the changes taking place in the Layer 9, to
illustrate the methods. For realistic use of gravity data however a simulation
of all the layers is necessary since filtering out effects for a single layer is
difficult due to the global nature of the gravity response. In addition we will
consider the total mass know. This will not be the case for Layer 9, but will
also require the simulation of all layers. The linear equations can then be
added corresponding to equation \eqref{eq:total_mass}.
\begin{equation}
\rmd M = N^{T} \rmd \theta_m.
\end{equation}
Here $\theta_m$ indicates the multiplicative variable of parameter $\theta$. We consider
all years with equal weights and standard deviation for the height $\sigma_h= 1
$ m and for gravity we consider $\sigma = 3 \, \mu \mathrm{Gal}$. The values of the
control parameters is non dimensional and relative to the original values. 
Table~\ref{tab:svd_values} confirm the analysis in section
\ref{sec:invariants} that measurement of the plume shape  give 2
approximate zero singular values. Gravity gives one, but adding the total mass as
a last constraint gives a determined system. In particular for the original case,
the plume shape helps in the two last singular values. This we attribute to the
smaller sensitivity of the gravity measurements to plume shapes, which is
important for determining part of the system. If we look at the singular vectors we
see that the smallest singular value of gravity and total mass system is
dominated by changes a combination of changes in porosity and permeability,
However this in the case total mass is include this included the space of the
lowest singular values of the plume shape.

\begin{table}
	\begin{center}
		\begin{tabular}{c | c c c c |c c c c}
			\toprule
			  & \multicolumn{4}{c|}{Original} & \multicolumn{4}{c}{Matched} \\
			\midrule
			P &  193 &  153 &    0 &    0  &  249 &  103 &    0 &    0\\ 
G &  556 &   32 &   27 &    0  &  427 &  139 &   30 &    0 \\ 
P G &  578 &  170 &   92 &    0 &  449 &  238 &  131 &    0\\ 
P M & 1421 &  192 &   63 &    0 & 1931 &  239 &   67 &    0\\ 
G M & 1445 &  470 &   29 &   13  & 1929 &  423 &  138 &   17\\ 
P G M & 1453 &  490 &  137 &   50 & 1932 &  448 &  232 &   75  

			\\
		    \bottomrule
		\end{tabular}
		\end{center}

	\caption{Table over the singular values of all time steps for original and match
		model.Here P, G, and M stand for plume, gravity and total mass perspectively. }
	\label{tab:svd_values}
\end{table}

\subsection{Sensitivity of topsurface data}
We now investigate the sensitivity with regard to changes in the top-surface. This
can be described by the relation
\begin{equation}
\rmd h_t = A_t \rmd z,
\end{equation}
where $\rmd h_t$ is the change in \co thickness from the matched model
corresponding to a
change in the topsurface $\rmd z$ at a given time $t$-. 
In the previous paper we showed that  $h+z$ was 
near constant, for situation near local equilibrium. 
We therefor also 
define modified matrices corresponding to mappings between parameters and
linear
combination of changes in \co thickness and top surface
\begin{equation}
\rmd h+\rmd z = A_{s,t} \rmd z \quad A_{s,t}=A_{t}+I
\end{equation}
$A_{s,t}$, can be seen as calculating the sensitivity with respect to the water-\co
interface
instead of the \co height. 
\begin{figure}[tbp]
	\centering
	
	\begin{tabular}{c c c}
		Original & Matched \\
		\includegraphics[width=0.45\textwidth]{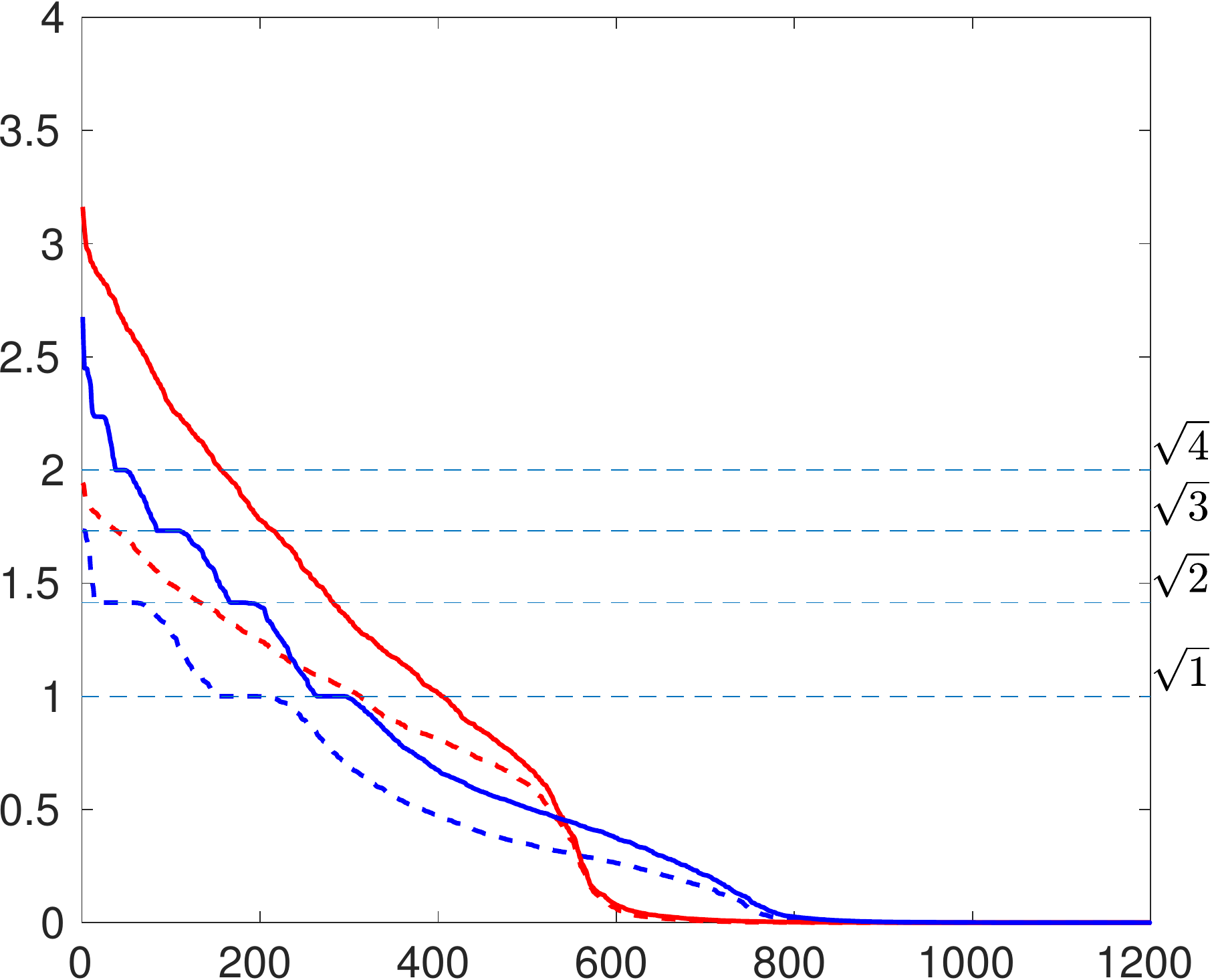} &
		\includegraphics[width=0.45\textwidth]{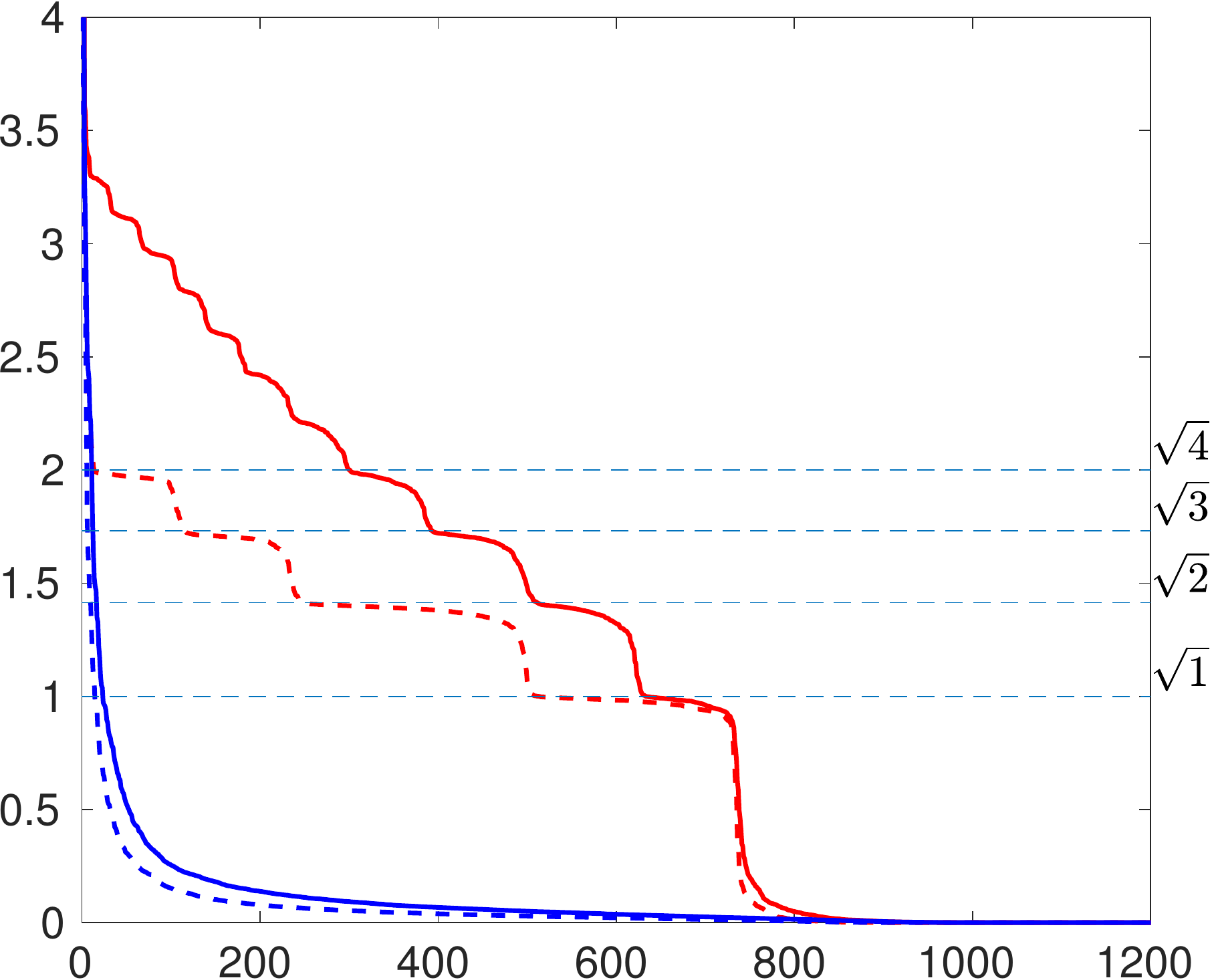}
	\end{tabular}
	\caption{The singular values of the full sensitivity. \textcolor{red}{Red color}
		signifies for $h$, i.e. plume observation while the \textcolor{blue}{blue color} is for $\rmd
		z$, interface observations. Full line is using all steps while the dashed line
		is using steps $1,5,9,12$. 
		}
	\label{fig:svd_time_plume}
\end{figure}

\begin{figure}[tbp]
	\centering
	\begin{tabular}{c c}
		Original & Matched \\		\includegraphics[width=0.45\textwidth]{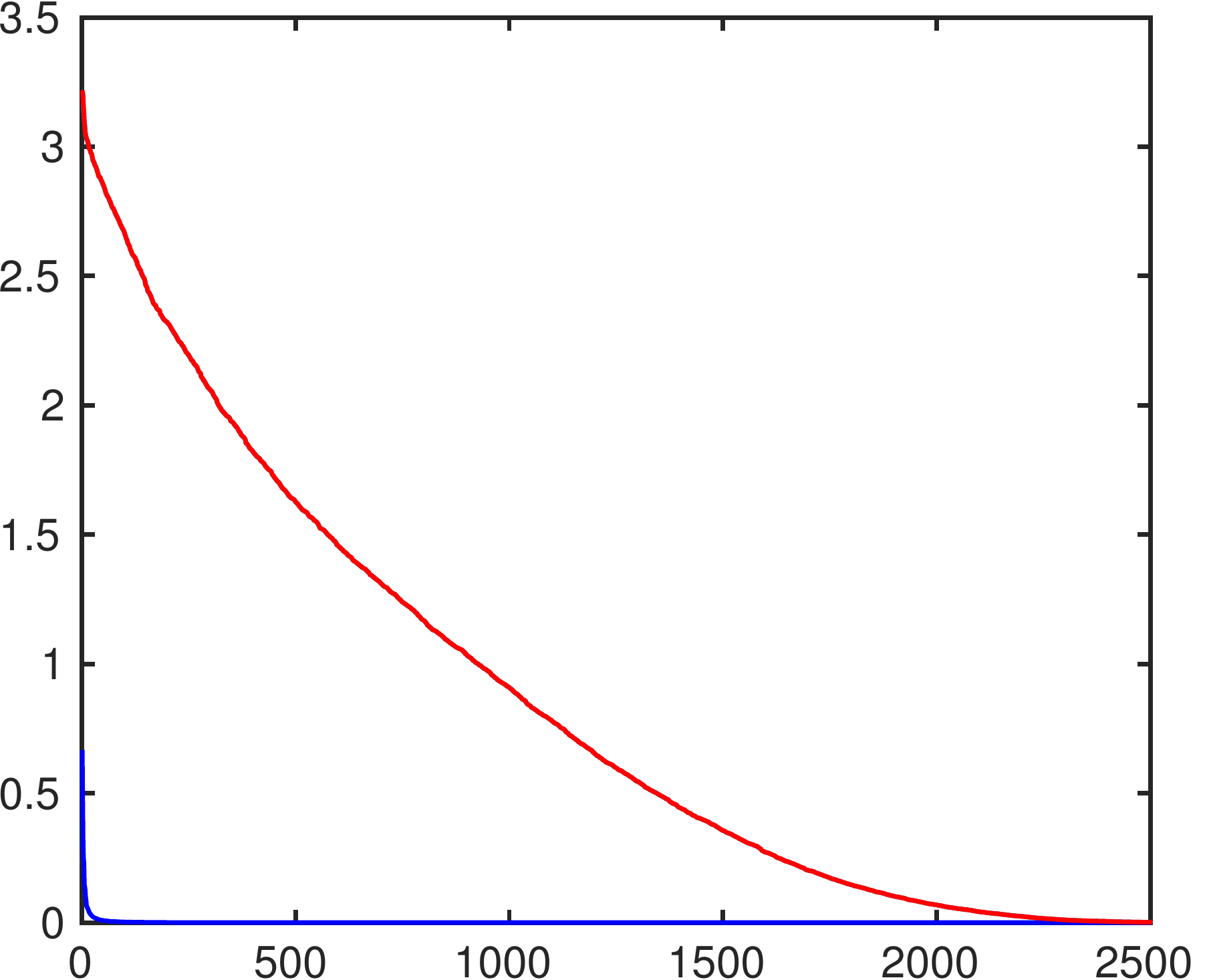} &
		\includegraphics[width=0.45\textwidth]{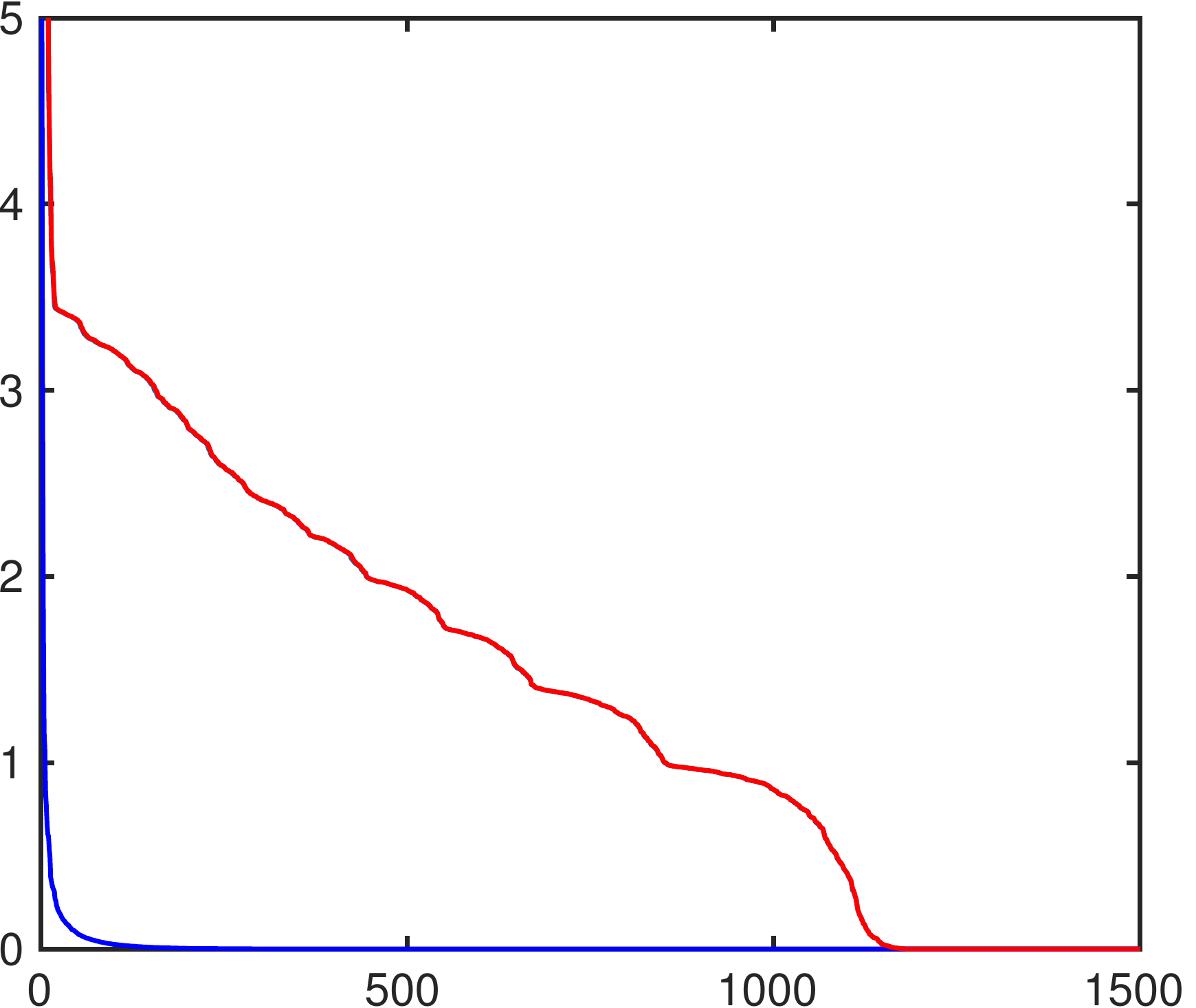}
	\end{tabular}
	\caption{The singular values of the full sensitivity. \textcolor{red}{Red color}
		signifies for plume observation while \textcolor{blue}{blue color} is gravity, $h$.}
	\label{fig:svd_time_gravity}
	
\end{figure}

Figure~\ref{fig:svd_time_plume} shows the singular for all time steps full line and
only including time steps of year $1$, $5$, $9$,and $12$, dashed line. The
\textcolor{red}{red} line is the sensitivites for plume height 
\textcolor{blue}{blue} is for the \co-water. We notice the much stronger
sensitivites of the match model for plume shape, while much less sensitivites
for the \co-water interface. This is since the \co-water interface
dynamics is only sensitive to the large scale features in the well segregated
model, in analogy with that a lake surface do not depend on the bottom shape. On
the other hand for the original model where gravity is much less important the
sensitivity to the interface of the plume shape is almost equal. In this case
the potential roughness of the top \co-water interface will not be smoothed by
the plume dynamics and topsurface altrations my be recognized in the  interface
dynamics.

If compare the sensitivities of the plume shape with the sensitivity of the
gravity measurement for the top surface, we see that there the gravity have a
fast decay, Figure~\ref{fig:svd_time_gravity}. There is hence very little part of the top surface changes which
have significant impact on the gravity response. This may be expected since
gravity only notice topsurface changes which contrite to significant global
changes in shape. Even more it must lead to significant changes in integrated
quantities, see equation \ref{eq:grav_changes}

\foreach \n in {9} {
\begin{figure}
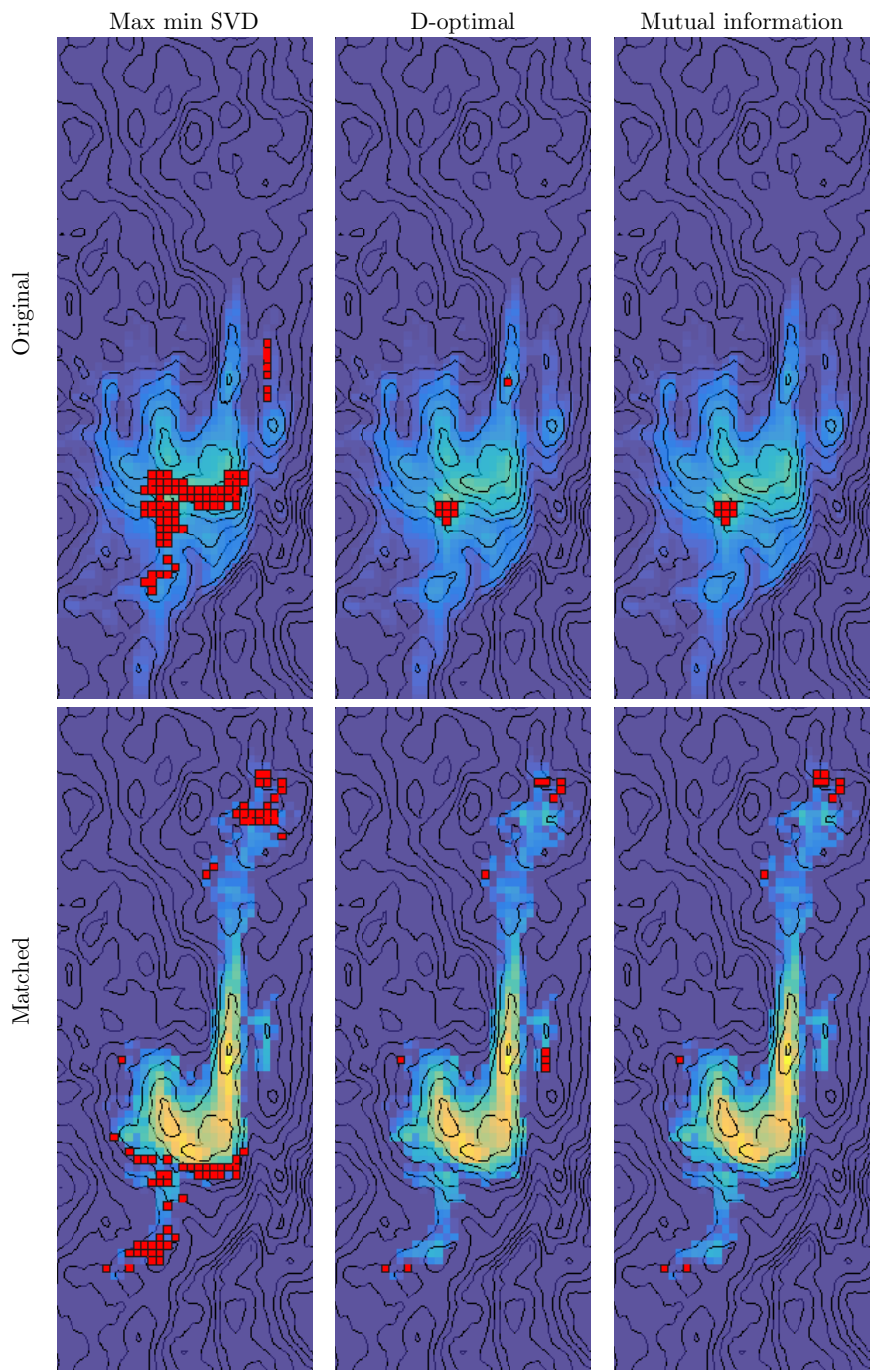
	
		\begin{center}
			\scalebox{0.9}{
				\begin{tabular}{l c c c}
					&Max min SVD & D-optimal  & Mutual information	\\
					\rotatebox{90}{\hspace{5cm} Original} &
					\includegraphics[clip=true, trim=170 40 170 70,width=0.3\textwidth]{./figs/opt_svd_step_\n_\casename_orginal.png} &
					\includegraphics[clip=true, trim=170 40 170 70,width=0.3\textwidth]{./figs/opt_dopt_step_\n_\casename_orginal.png} &
					\includegraphics[clip=true, trim=170 40 170 70,width=0.3\textwidth]{./figs/opt_mutual_step_\n_\casename_orginal.png}\\
					\rotatebox{90}{\hspace{5cm} Matched} 
					&\includegraphics[clip=true, trim=170 40 170 70,width=0.30\textwidth]{./figs/opt_svd_step_\n_\casename_optimum.png} &
					\includegraphics[clip=true, trim=170 40 170 70,width=0.30\textwidth]{./figs/opt_dopt_step_\n_\casename_optimum.png} &
					\includegraphics[clip=true, trim=170 40 170 70,width=0.30\textwidth]{./figs/opt_mutual_step_\n_\casename_optimum.png}
				\end{tabular}
			}
		\end{center}
	\caption{Comperaing different optimal design criterieas including timestep $1$,$5$,$9$ and $10$. The optimal points shown in \textcolor{red}{red} is for step $\n$}\label{fig:opt_comp}
\end{figure}
}
\foreach \n in {1} {%

		\begin{figure}
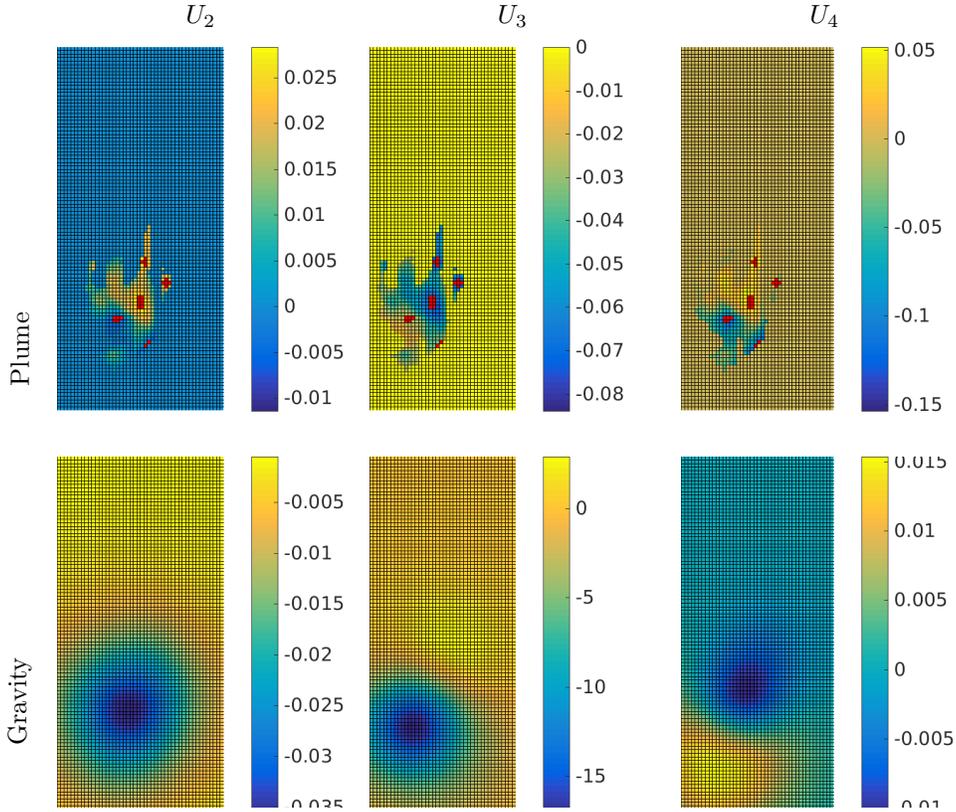

		\begin{center}
			\begin{tabular}{c c c c}
				& $U_2$ & $U_3$ & $U_4$\\
				\rotatebox{90}{	\quad \quad Plume }
				&
				\includegraphics[clip=true, trim=250 450 970 50,width=0.3\textwidth]{./figs/h_grav_param_hfac_\n_tstep9_\casename_orginal.png}
				&
				\includegraphics[clip=true, trim=670 450 550 50,width=0.3\textwidth]{./figs/h_grav_param_hfac_\n_tstep9_\casename_orginal.png}
				&
				\includegraphics[clip=true, trim=1070 450 150 50,width=0.3\textwidth]{./figs/h_grav_param_hfac_\n_tstep9_\casename_orginal.png}\\
				\rotatebox{90}{	\quad \quad Gravity }
				&
				\includegraphics[clip=true, trim=250 100 970 450,width=0.3\textwidth]{./figs/h_grav_param_hfac_\n_tstep9_\casename_orginal.png}
				&
				\includegraphics[clip=true, trim=670 100 550 450,width=0.3\textwidth]{./figs/h_grav_param_hfac_\n_tstep9_\casename_orginal.png}
				&
				\includegraphics[clip=true, trim=1070 100 150 450,width=0.3\textwidth]{./figs/h_grav_param_hfac_\n_tstep9_\casename_orginal.png}
			\end{tabular}
		\end{center}
		\caption{Optimal design for the original model with standard with the original assumption of the errors for gravity and plume data. The \textcolor{red}{red} squares show the optimal design usind D-optimal criteria with $20$ possible observation. The upper row show the response singular value corresponding to plume and the lower row for the gravity. The columns represent the second, third and forth singular value.}
		\label{fig:opt_grav_1_original}
		\end{figure}
			
}

\foreach \n in {10} {%
	\begin{figure}
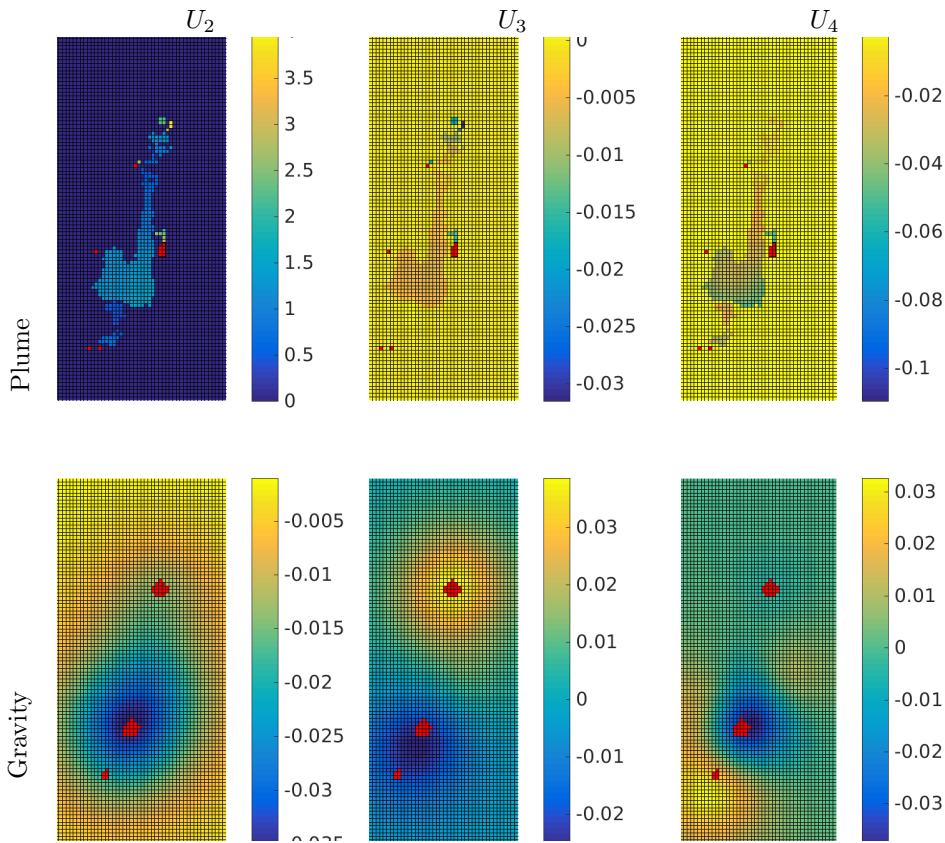

		\begin{center}
			\begin{tabular}{c c c c}
				& $U_2$ & $U_3$ & $U_4$\\
				\rotatebox{90}{	\quad \quad Plume }
				&
				\includegraphics[clip=true, trim=250 450 970 70,width=0.3\textwidth]{./figs/h_grav_param_hfac_\n_tstep9_\casename_optimum.png}
				&
				\includegraphics[clip=true, trim=670 450 550 70,width=0.3\textwidth]{./figs/h_grav_param_hfac_\n_tstep9_\casename_optimum.png}
				&
				\includegraphics[clip=true, trim=1070 450 150 70,width=0.3\textwidth]{./figs/h_grav_param_hfac_\n_tstep9_\casename_optimum.png}\\
				\rotatebox{90}{	\quad \quad Gravity }
				&
				\includegraphics[clip=true, trim=250 100 970 450,width=0.3\textwidth]{./figs/h_grav_param_hfac_\n_tstep9_\casename_optimum.png}
				&
				\includegraphics[clip=true, trim=670 100 550 450,width=0.3\textwidth]{./figs/h_grav_param_hfac_\n_tstep9_\casename_optimum.png}
				&
				\includegraphics[clip=true, trim=1070 100 150 450,width=0.3\textwidth]{./figs/h_grav_param_hfac_\n_tstep9_\casename_optimum.png}
			\end{tabular}
		\end{center}
		\caption{The same as Figure~\ref{fig:opt_grav_1_original}, but for the matched model and assuming standard deviation of gravity is $10$ times less.}
				\label{fig:opt_grav_10_matched}	
\end{figure}
}
\foreach \n in {1}{
		\foreach \cc in {optimum}{
  \begin{figure}
	 \begin{tabular}{c c c c}
	 & $U_2$ & $U_3$ & $U_4$\\
	 \rotatebox{90}{	\quad \quad Plume }
	 &
	 \includegraphics[clip=true, trim=250 450 970 50,width=0.3\textwidth]{figs/h_grav_param_hfac_strip\n_tstep9_\casename_\cc.png}
	 &
	 \includegraphics[clip=true, trim=670 450 550 50,width=0.3\textwidth]{figs/h_grav_param_hfac_strip\n_tstep9_\casename_\cc.png}
	 &
	 \includegraphics[clip=true, trim=1070 450 150 50,width=0.3\textwidth]{figs/h_grav_param_hfac_strip\n_tstep9_\casename_\cc.png}\\
	 \rotatebox{90}{	\quad \quad Gravity }
	 &
	 \includegraphics[clip=true, trim=250 100 970 450,width=0.3\textwidth]{figs/h_grav_param_hfac_strip\n_tstep9_\casename_\cc.png}
	 &
	 \includegraphics[clip=true, trim=670 100 550 450,width=0.3\textwidth]{figs/h_grav_param_hfac_strip\n_tstep9_\casename_\cc.png}
	 &
	 \includegraphics[clip=true, trim=1070 100 150 450,width=0.3\textwidth]{figs/h_grav_param_hfac_strip\n_tstep9_\casename_\cc.png}
	\end{tabular}
	\caption{The same as Figure~\ref{fig:opt_grav_10_matched} assuming observations of plume is in stripes. The standard deviations are the originally but the stripes are weighted with the number of cells.}
	\label{fig:opt_grav_1_stripe_matched}
  \end{figure}
}
}

\foreach \n in {1}{
	\foreach \cc in {optimum}{
		\begin{figure}
			\begin{tabular}{c c c c}
				& $U_2$ & $U_3$ & $U_4$\\
				\rotatebox{90}{	\quad \quad Plume }
				&
				\includegraphics[clip=true, trim=250 450 970 50,width=0.3\textwidth]{figs/h_grav_param_hfac_strip\n_tstep9_\casenamec_\cc.png}
				&
				\includegraphics[clip=true, trim=670 450 550 50,width=0.3\textwidth]{figs/h_grav_param_hfac_strip\n_tstep9_\casenamec_\cc.png}
				&
				\includegraphics[clip=true, trim=1070 450 150 50,width=0.3\textwidth]{figs/h_grav_param_hfac_strip\n_tstep9_\casenamec_\cc.png}\\
				\rotatebox{90}{	\quad \quad Gravity }
				&
				\includegraphics[clip=true, trim=250 85 970 450,width=0.3\textwidth]{figs/h_grav_param_hfac_strip\n_tstep9_\casenamec_\cc.png}
				&
				\includegraphics[clip=true, trim=670 85 550 450,width=0.3\textwidth]{figs/h_grav_param_hfac_strip\n_tstep9_\casenamec_\cc.png}
				&
				\includegraphics[clip=true, trim=1070 85 150 450,width=0.3\textwidth]{figs/h_grav_param_hfac_strip\n_tstep9_\casenamec_\cc.png}
			\end{tabular}
			\caption{The same as Figure~\ref{fig:opt_grav_10_matched} assuming observations of plume is in stripes. The standard deviations are the originally but the stripes are weighted with the number of cells.}
			\label{fig:opt_grav_1_stripe_matched_c}
		\end{figure}
	}
}

\begin{figure}
	\begin{center}
		\includegraphics[width=0.45\textwidth]{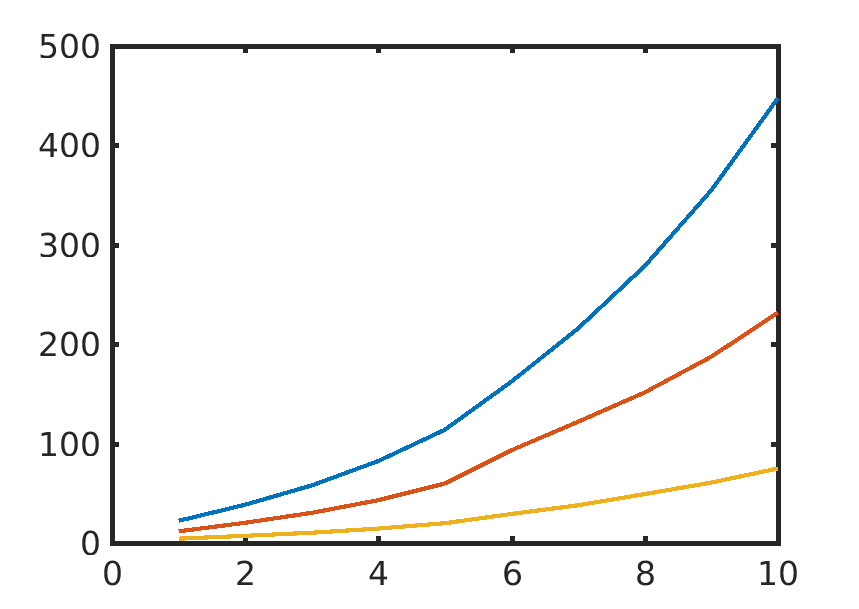}
		\includegraphics[width=0.45\textwidth]{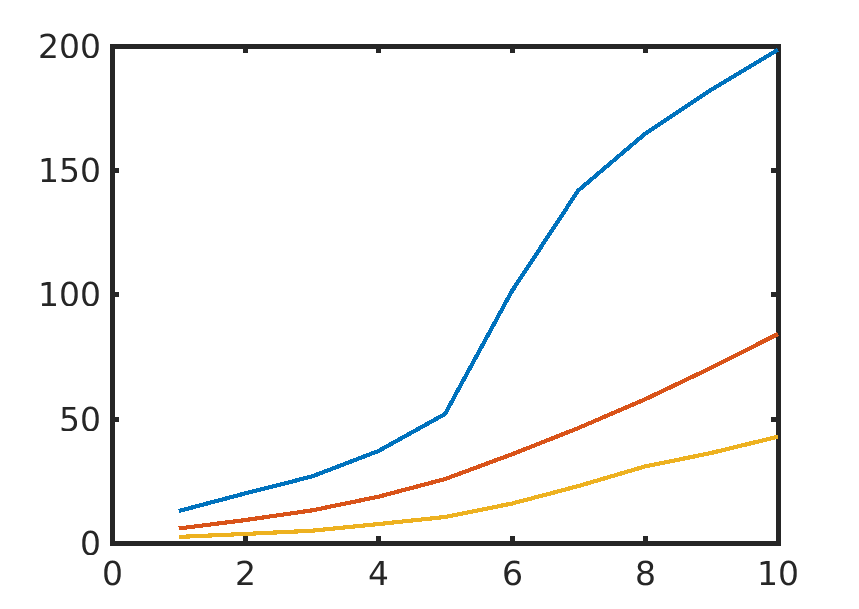}
	\end{center}
    \caption{The figure show the time-dependence of the accumulated svd of plume in blue, gravity in red and total mass in light brown with respect to the global parameters. The left is the using the direct observations. While the right show the same for using changes between time steps. 
    }\label{fig:svd_accum_full_diff}
\end{figure}

\section{Optimal design of experiment}
We now continue with the question of which of the parameters which make most
contributions to the observations. This depend on 
the cost and accuracy of different measurement technologies. For example in the case of seismic
 one have the choises between 2D, 3D with repeated survies or
stationary sensors. 3D is seen as the most cost efficient for deep aquifers with
in the context of oil exploration at an approximate cost of 10000\, \$/km$^2$. The cost of 2D seismic is 5000\, \$/km$^2$. However the cost to benefit evaluation in the context of oil extrapolation is not  directly applicable to \co monitoring.
The stationary case will in most cases
have better repeatability which may contribute to significant increase of
accuracy in estimation when a reservoir model is to be estimated. Gravity is
considered to be a cheaper for a given area. Often it is referred that the cost is reduced by a factor of 10, but to our knowledge a estimates taking into account the different information of the two services has not been published. The advantage is that it directly measures
model changes, but it have smaller sensitivity to
small spatial changes in the plume  which may be important for detecting leakage
pathways early. In the below we will only consider question of who to choose the
observation from a traditional optimal design of experiment perspective and not
take cost and accuracy of the different methods quantitatively into account by
scaling the importance of gravity with regard to plume shape. For a proper
evaluation in the case of plume shape it would have been necessary to include the
seismic inversion, since this will more directly be related the physical
devices which decide the cost of the measurements.
\subsection{Plume shape data}
We first look directly on the plume shape data and compare the
different methods of optimal design using the timesteps $1$,$5$,$9$ and $12$. We make the rather unrealistic assumption 
of been able to choose observations independently in different grid cells.
As expected the most important time steps is the last ones. The contributions from time step $9$ is shown in Figure~\ref{fig:opt_comp}. We see that all
optimal design criteria highlights the outline of the plume in the matched case. The effect is less pronounced for the original case. Another
important feature is that it divides the points used between the different
compartments. As discussed above the use of point data do not honor the way data is collected
and certainly not how the preprocessing step of seismic inversion is done.
A more fair use although simplified is to say that the each observation is either a slice in the x-direction or/and y-direction. We will consider this when including gravity data.

In all of these cases, we see that the different optimal design criteria give
approximately the same value. In terms of computational cost the ones based on
continuous optimization are most efficient, while the mutual information
based method give slightly more robust points form a qualitative viewpoint. The method based on
disciplined convex programing seemed more stable for large systems. This may
be due to the slightly simpler D-optimal criteria. The interior
point based algorithm is much simpler to extend to a cost based optimization However
convexity and a guaranty to find the global optimum can not be known a priory. For the rest of the paper we will
stick the the computational attractive D-optimal criteria.

\subsection{Optimal design for gravity and plume shape data}
We know look at the effect of combining plume shape and gravity data
together. To form a complete system for the global parameters and to be more
realistic for a full system, we restrict us to the case where the total
mass is known accurately. We consider two different scenarios. One where we us
the
unchanged sensitivities and one where we multiply the gravity sensitivity with a
factor $10$. Figure~\ref{fig:opt_grav_1_original} and
\ref{fig:opt_grav_10_matched} show the results for 
the original model with factor $1$ and the matched model factor $10$ respectively,
using time step 9. First we notice the the placement of the 
gravity observations is more prone centered than for the plume data. This we
attribute to the global smooth character of the gravity response.
Secondly we observe that in the case where it is favorable to have many gravity
observation, Figure~\ref{fig:opt_grav_10_matched}, the gravity observations is
placed to be sensitive to the multipole expansion of the gravity response. We
also see that even with this low weight of the plume data it is still favorable
to keep some of this observation. In particular the observations on the other
side of the main spill point is favored. The colors of the figure give the
strength for different singular value in each columns and the two rows response singular vector
for the plume upper row and gravity lower row. We see that the two first singular values  in the high gravity sensitivity case, Figure ~\ref{fig:opt_grav_10_matched}, are dominated
monopole and the dipole contribution respectively.

The assumption of being able to choose the estimates of the plume shape freely
in space is not realistic both with respect to who seismic data is 
collected offshore and also with the preprocessing step needed
estimating the plume shape from raw seismic wavefields. To make the optimal
design somewhat more realistic without introducing seismic inversion we
restrict observations of the plume to be either in horizontal strips or
in vertical strips. We weight the observations with the number of points in a
strips. This would correspond to letting the cost of one strip be proportional to
the length. Figure ~\ref{fig:opt_grav_1_stripe_matched} show the result for the
matched case with equal original weight. This is still a case which
slightly favoring gravitational measurements, but we see that the two slices of
the plume are chosen in the region where the plume is thickest.

\subsection{Effect of diffential measurements}
For many measurements more accuracy can be obtained for differences than for
a given value. This is particular the case for technologies with high
repeatability and small drift. In the case of monitoring \co this will in
particular be the case for certain types of gravity measurements. Passive
monitoring systems for seismic waves where position of the hydrophones is the
same all the time will also potentially benefit from processing which can utilize
the this feature.
Given our linearized model the advantages of large repeatability
can be evaluated using
\begin{equation}
\rmd (y_{i+1}- y_{i}) = (A_{i+1}-A_{i})\rmd \theta
\end{equation} 
If the errors associated with measurements of $y_{i+1}- y_{i}) $ is
significantly less than for the original quantities large gains can be achieved,
if the measurements are linear independent with 
regard to the parameters. That is $A_{i+1}$ is significantly different from
$A_{i}$. Figure~\ref{fig:svd_accum_full_diff} show the evolusion of the svd of
the accumulated system using observation of absolute changes left and
differential changes right. Given that the  measurements system have more
	than $4$ times lower standard deviation for differential measurement, that will give more information.

\subsection{Effect of resolution}
The framework may become computational challenging when the resolution is
increased. Figure
~\ref{fig:opt_grav_1_stripe_matched_c} show the same results as in
Figure~\ref{fig:opt_grav_1_stripe_matched} but with a coarser resolution.
We see that the optimal design is qualitatively the same. We have seen the same feature
for most of the cases we have run. Some discrepancies is found when resolution changes the behavoir of the plume. Most prominently when the upward flowing feature is underestimated due to low resolution.

\section{Discussion}
When considering an optimal monitoring strategy one has to consider the
cost of a measurement, how much it contributes to the knowledge of the system and
how valuable the information is. The cost is most often known, but
quantifying how well one can estimate the system given a set of measurements is
difficult. This requires a model of the system and the measurement methods. 
For accurate estimation one also need a good estimation of the errors and covariance of
the error (even neglecting the difficulties of nonlinearities.) This may include
pure stochastic errors of the measurement process, but more often it is
associated with uncertainties in parameters of the measurement process (i.e.
inverse modeling errors) which often depend on unknown geology. Some of this
parameters have very large variations like permeability, and some have less
variation like density of stone. However how important the variation is for the
system is highly dependent on the system. A key point is the covariance of the errors of a measurement and the linearly independence of them. 
Roughly speeking correlation is good for linearly independent measurements but bad for linearly dependent measurements.
To evaluate the value of the
information one in addition have to connect the information with a value of a
decision process. The latter is a difficult question in itself for \co storage,
which only have indirect value and the responsibility of may be shared between different stakeholders.

Instead of starting from the measurement process we have in this paper
investigated the question from the simplest simulation model which has been
demonstrated to simulate \co plume dynamics accurately in certain situation,
most prominently the Sleipner L9 case. The motivation has been to look at the
sensitivity of this simple model and connect this to observation, although in
an indirect way in the case of plume shape. With this simplification we have
been able to calculate full linear models from the parameters of the nonlinear model to the
observations. This has made it possible to investigate both how knowledge of
the plume dynamics and gravity response relate to the model parameters. 
In particular we could exploit the power of the SVD compositions. In addition, it made it possible to
investigate the monitoring problem efficiently in the setting of optimal design.
We consider this to be a starting point for optimization of the monitoring strategy with respect to
accuracy constraints. Furthermore our work should be extended for
optimizing the monitoring framework so that it is robust with respect to the
assumed uncertainties. As we have demonstrated in the Sleipner L9 case the naive
answer to the optimal design question is significantly different between the
original assumed model and what we today believe is the reality. A monitoring
policy should therefor be robust and with respect to this initial large uncertainty.

In addition to the theoretical difficulties of the problem in itself we
have the question: Is the problem computational feasible?
We have here not included the seismic inversion process which is a large
computation challenge in it self, we believe could benefit
form a direct coupling with the flow modeling in particular if the inversion is
targeted to changes. We have used a VE based simulation model with
and restricted the uncertainties of the model to rather few parameters. Extending
this model will contribute to increased computation cost, in particular if full
3D model is required. However we believe that extensions to layered models will
be sufficient form many cases. Also the assumption of uniform uncertainty of
permeability will in many cases be to simplistic. However both of this problems
contribute to increasing the computation of the above frame work, they do grow
approximately linearly due to the use of adjoint bases sensitivities. When
combined with full linearization of the system which also grow linearly with
the number of observation it may be quite restrictive. The linearization will
most often lead to dens
systems. This may be challenging for a strait forward full evaluation of the SVD
decomposition. However the observations
is independent of the grid gridding and may be restricted. SVD analysis can also
give valuable information about particular subspaces.
Coarse grids in combination with VE based models is also likely to give good
result due to the strong parabolic term in this equation. 
In addition to the challenge of including nonlinear and robust optimization in a
statistical sense, the critical
part is if the problem of optimal design can be done if the model complexity is
larger. We believe this
is less computation demanding than the full SVD calculations. In addition one
would in most cases not seek full free optimization like in 
Figure~\ref{fig:opt_grav_10_matched} and \ref{fig:opt_grav_1_original} but
rather restrict for example the possible configurations to strips and a few
points like in \ref{fig:opt_grav_1_stripe_matched}. This reduce the
computational cost significantly and make the framework valuable to
evaluate cost versus benefit at least for restricted combination of
measurements.

\section{Conclusion}
In the setting of \co monitoring we have used flow simulations with adjoint
based sensitives and shown how efficient the different monitored variables
are to estimate a simulation model. In particular we have by using an
efficient VE-based simulation model shown that different physical scenarios
thought to be plausible for the Sleipner Layer 9 have different qualitatively with respect to the sensitivity to 
monitored quantities. We showed that the main
parameters of our model could be estimated if gravity response and total mass is
known. Plume shape and total mass is, however not sufficient. But plume-shape
gave significant additional benefit even not sufficient in it self. 

We also used the framework to evaluate the estimated quantities in a setting of
optimal design. Although not giving quantitatively estimate of efficient
monitoring strategies this gave insight into which part of the estimated
quantities are most important for estimating the flow model. With the simplified
model for accuracy the optimal design gave high weights to the plume outline
while the important part of the gravity response was the main multipoles. As
gravity response is global and effectively isolated to the change in \co volume
times the density difference to water, it seems like an ideal candidate to
robustly monitor the main part of the \co dynamics. Little assumption of the
overburden is also needed, although separating the gravity changes of the \co to
other sources may be difficult without a proper simulation model which can give
strong restrictions on how gravity should change both in time and space.
Our results indicate that gravity measurements should be a valuable part of a
monitoring program details of the plume shape give important complementary
information and a combination is likely to be the most efficient. In a scenario
where cost is introduced our results indicate that limited accurate observations
will give the most cost efficient and robust monitoring strategy. But more
in-dept study using realistic cost to accuracy, including models for seismic
inversion is needed to establish this further. When a reliable dynamic model can
be formulated the result indicate that observation with high accuracy on
differential measurements will be favorable. 

In this paper we discussed the use of optimal design in the simplified setting
of Sleipner Layer 9. In more difficult cases where strong restrictions of the
injectivity, possibility of activation of near by faults measurements related
to pressure would probably be favorable. The possible global measurements are in
such case uplift or possible micro-seismic activity. Both will most likely favor
stationary passive systems. Uplift which my be readily included in our framework
will be most sensitive to differential measurements with high accuracy, since this
can be attributed directly to pressure changes in a similarly to gravity, this
is contrary to seismic and electromagnetic measurements where where only measure
changes in the media indirectly. Mico seismics also look at a source to the
signal directly, but here the source is more indirectly assosiated with the \co
injection.

We believe the computational tools and methods used here in the setting of \co
storage will contribute to finding cost efficient monitoring programs for \co
storage.

\section*{Acknowledgements}

This work was funded in part by the Research Council of Norway through grant
no.\ 243729
(Simulation and optimization of large-scale, aquifer-wide \co injection in the
North Sea).

Statoil and the Sleipner License are acknowledged for provision of the Sleipner
2010
Reference dataset. Any conclusions in this paper concerning the Sleipner field
are the
authors' own opinions and do not necessarily represent the views of Statoil.

\bibliographystyle{elsarticle-num} 
\bibliography{refs_tccs_clean.bib}
\end{document}